

\input harvmac

 \def \k1 {{1\over
k}} \def \bh { {\bar h} } \def \ov { \over }

\def \ra {\rightarrow}

\def \a {\alpha}
\def \b {\beta}

\def \Tr {{\ \rm Tr \ }}

\def \ln {{\rm \ ln \  }}
\def \det {{\ \rm det \ }}

\def \th {{\rm tanh  }}
\def \l {\lambda}
\def \1p {{1\over  \pi }}
\def \2p {{{1\over  2\pi }}}
\def \4p {{ {1\over 4 \pi }}}
\def \8p {{{1\over 8 \pi }}}
\def \P^* { P^{\dag } }
\def \p {\phi}

\def \ga {\alpha}
\def \gg {\gamma}
\def \gb {\beta}

\def \m {\mu }
\def \n {\nu}

\def\g {\gamma}
\def \r {\rho}
\def \k {\kappa }

\def \o {\omega}
\def \s {\sigma}
\def \t {\theta}

\def \fourth {{\textstyle{1\over 4}}}

\def \e#1 {{{\rm e}^{#1}}}

\def \eq#1 {\eqno {(#1)}}
\def \sm {sigma model\ }

\def \bd  {{ \bar \del }}

\def \E {{ \tilde E}}

\def \bd  { \bar \del }

\def \ov {\over }

\def \A  { {\bar A} }
\def \tth {\tilde h}

\def \H {{\cal H}}
\def \o {\omega}
\def \gl {\lambda}
\def \gp {\phi}
\def \p {\phi}

\def \s {\sigma}
\def\gg {\gamma}
\def \gr {\rho}
\def \r {\rho}

\def \l {\lambda}
\def \m {\mu}
\def \g {\gamma}
\def \n {\nu}

\def \fourth {{1\over 4}}

\def \e#1 {{{\rm e}^{#1}}}

\def \hg {{\hat g}}

\def \H {{\cal H}}

\def \tg {{\tilde g}}

\def\np {  Nucl. Phys. }
\def \pl { Phys. Lett. }
\def \mpl { Mod. Phys. Lett. }
\def \prl { Phys. Rev. Lett. }
\def \pr  { Phys. Rev. }
\def \ap  { Ann. Phys. }
\def \cmp { Commun. Math. Phys. }
\def \ijmp { Int. J. Mod. Phys. }
\baselineskip10pt
\Title{\vbox
{\baselineskip8pt{\hbox{CERN-TH.6872/93}}{\hbox{RI-150-93}}{\hbox{hep-th/9304155}} }}
{\vbox{\centerline { Heterotic string  solutions }\vskip2pt
 \centerline{ and coset  conformal field theories   }
}}
\centerline { Amit Giveon\footnote{$^*$}{e-mail address:
giveon@hujivms.bitnet},
  \ Eliezer  Rabinovici\footnote{$^{**}$}{e-mail address:
eliezer@hujivms.bitnet} }

\centerline {\it  Racah Institute of Physics, The Hebrew University}
\centerline {\it
 Jerusalem 91904, Israel}
\medskip
\centerline { and}
\medskip
\centerline{   A.A. Tseytlin\footnote{$^{***}$}{\baselineskip8pt
On leave from Blackett Laboratory, Imperial College, London SW7 2BZ, U.K.
and P.N. Lebedev Physics
Institute, Moscow, Russia.  $ \  $
e-mail address: tseytlin@surya3.cern.ch and tseytlin@ic.ac.uk} }

\centerline {\it Theory Division, CERN}
\centerline {\it
CH-1211 Geneva 23, Switzerland}
\medskip
\centerline {\bf Abstract}
\medskip
\baselineskip7pt
\noindent
We  discuss   solutions of the heterotic string theory  which are
analogous to bosonic and superstring backgrounds  related   to
coset conformal field theories.  A  class of  exact `left-right
symmetric' solutions is obtained by supplementing
the metric, antisymmetric tensor and dilaton of the
superstring solutions by the gauge field background equal
to the generalised Lorentz connection with
torsion. As in the superstring case, these backgrounds are
$\a'$-independent, i.e.  have `semiclassical'  form.
The   corresponding  heterotic string sigma model  is obtained  from
the combination of  the (1,0) supersymmetric gauged WZNW  action with
the  action of internal fermions  coupled to the target space
gauge field.  The  pure (1,0) supersymmetric gauged WZNW  theory
is anomalous and does not describe a consistent heterotic string solution.
We also find (to the order $\alpha'^3$)  a two-dimensional perturbative
heterotic string solution  with  the trivial gauge field background.
To  the leading  order in $\alpha'$  it coincides with the  known
$SL(2,R)/U(1)$ bosonic or superstring solutions. This solution
does not correspond  to a `heterotic' combination of the  left
superstring and right bosonic $L_0$-operators   at the conformal
field theory level.  Some duality properties of the heterotic
string solutions are  studied.

\bigskip
\noindent
{CERN-TH.6872/93}

\noindent
 {April 1993}
\Date { }

\noblackbox
\baselineskip 20pt plus 2pt minus 2pt

\lref \bep {C. Becchi and O. Piguet, \np B315(1989)153. }
\lref \nov {S.P. Novikov,  Sov. Math. Dokl. 37(1982)3. }
\lref \fuch { J. Fuchs, \np B286(1987)455 and  B318(1989)631. }
\lref \sus { R. Rohm,  \pr D32(1985)2845.}
\lref \suss {   H.W. Braden, \pr D33(1986)2411.}
\lref   \red { A.N. Redlich and H.J. Schnitzer, \pl B167(1986)315 and
B193(1987)536(E);  A. Ceresole,
 A. Lerda, P. Pizzochecco
 and
P. van Nieuwenhuizen, \pl
 B189(1987)34.}
 \lref \div  { P. Di Vecchia,  V. Knizhnik, J. Peterson and P. Rossi, \np
B253(1985)701.}

\lref \schn {  H. Schnitzer, \np B324(1989)412.  }

\lref \all { R.W. Allen, I. Jack and D.R.T. Jones, Z. Phys. C41(1988)323. }

\lref \nem {D. Nemeschansky and S. Yankielowicz, \prl 54(1985)620; 54(1985)1736
(E).}

\lref \bep { C. Becchi and O. Piguet, \np B315(1989)153. }

\lref \ks {Y. Kazama and H. Suzuki, \np B321(1989)232; \pl B216(1989)112.}

\lref \mor {A.Yu. Morozov, A.M. Perelomov, A.A. Rosly, M.A. Shifman and A.V.
Turbiner, \ijmp
A5(1990)803.}

 \lref \tur {A.V. Turbiner, \cmp 118(1988)467;  M.A. Shifman and A.V. Turbiner,
\cmp 126(1989)347;
M.A. Shifman, \ijmp A4(1989)2897.}

\lref \hal { M.B. Halpern and E.B. Kiritsis, \mpl A4(1989)1373; A4(1989)1797
(E).}

\lref \haly {M.B. Halpern and   J.P. Yamron,  Nucl.Phys.B332(1990)411;
Nucl.Phys.
B351(1991)333.}
\lref \halp { M.B. Halpern, E.B. Kiritsis, N.A. Obers, M. Porrati and J.P.
Yamron,
\ijmp A5(1990)2275;
  A.Yu. Morozov,  M.A. Shifman and A.V. Turbiner, \ijmp
A5(1990)2953;
A. Giveon, M.B. Halpern, E.B. Kiritsis and  N.A. Obers,
\np B357(1991)655.}
\lref \bpz {A.A. Belavin, A.M. Polyakov and A.B. Zamolodchikov, \np
B241(1984)333. }
\lref \efr {     S. Elitzur, A. Forge and E. Rabinovici, \np B359 (1991)
581;
 G. Mandal, A. Sengupta and S. Wadia, Mod. Phys. Lett. A6(1991)1685. }
\lref \sak {K. Sakai, Kyoto preprint, KUNS-1141-1992. }
\lref \ver {H. Verlinde, \np B337(1990)652.}
\lref \gwz {      K. Bardakci, E. Rabinovici and
B. S\"aring, \np B299(1988)157;
 K. Gawedzki and A. Kupiainen, \pl B215(1988)119;
\np B320(1989)625. }

\lref \sen {A. Sen, preprint TIFR-TH-92-57. }

\lref \bcr {K. Bardakci, M. Crescimanno and E. Rabinovici, \np
B344(1990)344. }
\lref \Jack {I. Jack, D.R.T.  Jones and J. Panvel,  \np B393(1993)95. }
\lref \zam  { Al. B. Zamolodchikov, preprint ITEP 87-89. }

\lref \hor {J. Horne and G. Horowitz, \np B368(1992)444. }
\lref \tse { A.A. Tseytlin, \pl B264(1991)311. }
\lref \gwzw  { P. Di Vecchia and P. Rossi, \pl  B140(1984)344;
 P. Di Vecchia, B. Durhuus  and J. Petersen, \pl  B144(1984)245.}
\lref \oal { O. Alvarez, \np B238(1984)61. }

\lref \ishi { N. Ishibashi, M.  Li and A. Steif, \prl 67(1991)3336. }
\lref  \kumar  { M. Ro\v cek and E. Verlinde, \np B373(1992)630; A. Kumar,
preprint CERN-TH.6530/92;
 S. Hussan and A. Sen,  preprint  TIFR-TH-92-61;  D. Gershon,
preprint TAUP-2005-92; X. de la Ossa and F. Quevedo, preprint NEIP92-004; E.
Kiritsis, preprint LPTENS-92-29. }

\lref \rocver { A. Giveon and M. Ro\v cek, \np B380(1992)128. }
\lref \frts {E.S. Fradkin and A.A. Tseytlin, \np B261(1985)1. }
\lref \mplt {A.A. Tseytlin, \mpl A6(1991)1721. }
\lref\bn {I. Bars and D. Nemeschansky, \np B348(1991)89.}
\lref \shif { M.A. Shifman, \np B352(1991)87.}
\lref\wittt { E. Witten, \cmp 121(1989)351; G. Moore and N. Seiberg, \pl
B220(1989)422.} \lref \chernsim { E. Guadagnini, M. Martellini and M.
Mintchev, \np B330(1990)575;
L. Alvarez-Gaume, J. Labastida and A. Ramallo, \np B354(1990)103;
G. Giavarini, C.P. Martin and F. Ruiz Ruiz, \np B381(1992)222; preprint
LPTHE-92-42.}
\lref \shifley { H. Leutwyler and M.A. Shifman, \ijmp A7(1992)795. }
\lref \polwig { A.M. Polyakov and P.B. Wiegman, \pl B131(1984)121; \pl
B141(1984)223.  }
\lref \polles { A. Polyakov, in: {\it Fields, Strings and Critical Phenomena},
  Proc. of Les Houches 1988,  eds.  E. Brezin and J. Zinn-Justin
(North-Holland,1990).   }
\lref \kutas {
D. Kutasov, \pl B233(1989)369.} \lref \karabali { D. Karabali, Q-Han Park, H.J.
Schnitzer and
Z. Yang, \pl B216(1989)307;  D. Karabali and H.J. Schnitzer, \np B329(1990)649.
}
\lref \ginq {P. Ginsparg and F. Quevedo,  \np B385(1992)527. }
\lref \gko  {K. Bardakci and M.B. Halpern, \pr D3(1971)2493;
 M.B. Halpern, \pr D4(1971)2398;   P. Goddard,
A. Kent and D. Olive, \pl B152(1985)88; \cmp 103(1986)303;  V. Kac and I.
Todorov, \cmp
102(1985)337.  }
\lref \dvv  { R. Dijkgraaf, H. Verlinde and E. Verlinde, \np B371(1992)269. }
\lref \kniz {  V. Knizhnik and A. Zamolodchikov, \np B247(1984)83. }

\lref \witt { E. Witten, \cmp 92(1984)455.}
\lref \wit { E. Witten, \pr D44(1991)314.}
\lref \anton { I. Antoniadis, C. Bachas, J. Ellis and D.V. Nanopoulos, \pl B211
(1988)393.}
\lref \bsfet {I. Bars and  K. Sfetsos, \pr D46(1992)4510; \pl B301(1993)183;
  K. Sfetsos,  preprint USC-92/HEP-S1 (1992).}

\lref \ts  {A.A.Tseytlin, \pl B268(1991)175. }
\lref \kir {{E. Kiritsis, \mpl A6(1991)2871.} }

\lref \shts {A.S. Schwarz and A.A. Tseytlin,  preprint Imperial/TP/92-93/01
(1992). }
\lref \bush { T.H. Buscher, \pl B201(1988)466.  }

\lref \plwave { D. Amati and C. Klim\v cik, \pl B219(1989)443; G. Horowitz and
A. Steif, \prl 64(1990)260.}

\lref\bsft { I. Bars, preprint USC-91-HEP-B3. }

\lref\bs { I. Bars and K. Sfetsos,  preprint USC-93/HEP-B1 (1993). }
\lref\bst { I. Bars,  K. Sfetsos and A.A. Tseytlin, unpublished. }
\lref\bb  { I. Bars, \np B334(1990)125. }
\lref \tsw  { A.A. Tseytlin, preprint Imperial/TP/92-93/10 (1992).}
\lref \ger { A. Gerasimov, A. Morozov, M. Olshanetsky, A. Marshakov and S.
Shatashvili, \ijmp
A5(1990)2495. }
\lref \sch { K. Schoutens, A. Sevrin and P. van Nieuwenhuizen,
in: Proc. of the Stony Brook Conference {\it `Strings and Symmetries 1991'},
p.558  (World
Scientific, Singapore, 1992).} \lref \boer { J. de Boer and J. Goeree, Utrecht
preprint THU-92/33. }
\lref \dev { C. Destri and H.J. De Vega, \pl B208(1988)255. }
\lref  \noj {S. Nojiri, \pl B271(1992)41. }
\lref \swz  { E. Witten, \np B371(1992)191;
T. Nakatsu, Progr. Theor. Phys. 87(1992)795. }
 \lref \bsf { I. Bars and K. Sfetsos, \pl B277(1992)269. }

\lref \br {A. Barut and R. Raczka, ``Theory of Group Representations and
Applications", p.120
 (PWN, Warszawa 1980). }
\lref \jjmo {I. Jack, D.R.T. Jones, N.Mohammedi and H. Osborn, \np
B332(1990)359;
C.M. Hull and B. Spence, \pl B232(1989)204. }

\lref \tttt { A.A. Tseytlin,  preprint Imperial/TP/92-93/7 (1992), \pr
D47(1993) no.8.}
\lref \per { V. Novikov, M. Shifman, A. Vainshtein and V. Zakharov, \pl
B139(1984)389;
A. Morozov, A. Perelomov and M. Shifman, \np B248(1984)279;
M.C. Prati and A.M. Perelomov, \np B258(1985)647. }
\lref \alv {
L. Alvarez-Gaum\'e,   D. Freedman and S. Mukhi, \ap 134(1981)85;
L. Alvarez-Gaum\'e, \np B184(1981)180. }
\lref \gr {  M.T. Grisaru, A. van de Ven and D. Zanon, \np B277(1986)409. }
\lref \call {C.G. Callan, D. Friedan, E. Martinec and M.J. Perry,
Nucl. Phys.B262(1985)593; A. Sen, Phys. Rev. D32(1985)316; \prl 55(1985)1846.}
\lref \hul {C.M. Hull, \pl B167(1986)51; Nucl. Phys. B267(1986)266.}
\lref \het { Y. Kikuchi, C. Marzban and Y.J. Ng, \pl B176(1986)57;
Y. Cai and C.A. Nunez, \np B287(1987)279;
Y. Kikuchi and  C. Marzban, \pr D35(1987)1400.}
\lref \met {R.R. Metsaev and A.A. Tseytlin, \pl B185(1987)52.}
\lref \gro {D.J. Gross and J.H. Sloan, \np B291(1987)41. }
\lref  \foa { A.P. Foakes, N. Mohammedi and D.A. Ross, \pl B206(1988)57;
\np B310(1988)335.}
\lref \hw {C.M. Hull and E. Witten, \pl B160(1985)398.}
\lref \hh { D.J. Gross, J.A. Harvey, E. Martinec and R. Rohm, \np B256(1985)253
and  B267(1986)75.}
\def \ho { {\hat \o } }
\lref \chsw { P. Candelas, G. Horowitz, G. Strominger and E. Witten, \np
B258(1985)46. }
\lref \aat { A.A. Tseytlin, preprint CERN-TH.6804/93.}
\lref \at {A.A. Tseytlin, preprint CERN-TH.6820/93. }
\lref \aps { S. De Alwis, J. Polchinski and R. Schimmrigk, \pl B218(1989)449. }
\lref \gny  { M.D. McGuigan, C.R. Nappi and S.A. Yost, \np B375(1992)421. }
\lref \py  {  M.J.  Perry and E. Teo, preprint DAMTP R93/1 (1993); P. Yi,
preprint CALT-68-1852
(1993). }
\lref \tv {   A.A. Tseytlin and C. Vafa, \np B372(1992)443.}
\lref \hull {C.M. Hull and B. Spence, \np B345(1990)493. }
\lref \gep { D. Gepner, \pl B199(1987)370; \np B296(1988)757.}
\lref \gat {S.J. Gates, S.V. Ketov, S.M. Kuzenko and O.A. Soloviev, \np
B362(1991)199.}
\lref \ell {U. Ellwanger, J. Fuchs and M.G. Schmidt, \pl B203(1988)244; \np
B314(1989)175. }
\lref \ket { S.V. Ketov and O.A. Soloviev, \pl B232(1989)75; \ijmp
A6(1991)2971. }
\lref \GK {A. Giveon and E. Kiritsis,  preprint CERN-TH.6816/93,
RIP-149-93.}
\lref \GR {A. Giveon and M. Ro\v{c}ek, Nucl. Phys. B380(1992)128.}
\lref \giv {  A. Giveon, Mod. Phys. Lett. A6(1991)2843.}
\lref \GRV {A. Giveon, E. Rabinovici and G. Veneziano, Nucl. Phys.
B322(1989)167;
A. Shapere and F. Wilczek, \np B320(1989)669.}
\lref \GMR {A. Giveon, N. Malkin and E. Rabinovici, Phys. Lett. B238(1990)57.}
\lref \GS  {  A. Giveon and D.-J. Smit. Nucl. Phys.  B349(1991)168.}
\lref  \GP   {  A. Giveon and A. Pasquinucci, Phys. Lett. B294(1992)162.}
\lref \GG  { I.C. Halliday, E. Rabinovici, A. Schwimmer and M. Chanowitz, \np
B268(1986)413. }
\lref \Ve    { K.M. Meissner and G. Veneziano, \pl B267(1991)33;
M. Gasperini, J. Maharana and G. Veneziano, \pl B272(1991)277;
A. Sen,  \pl  B271(1991)295.}

\newsec {Introduction }
Finding exact (all orders in $\a'$) solutions of string theory is a complicated
problem. Not only  do  the  string  equations contain terms of all orders in
the number
of derivatives, but also   the  explicit  form  of these equations
(`$\b$-functions')
or  the corresponding effective action is not known explicitly.
A possible strategy to find exact solutions is to first  determine the leading
order form of it
and then identify a   conformal field theory  which  generalises it to all
orders.
This program can be implemented  for   a large class of solutions
corresponding to
coset conformal field theories \gko\ which have a  Lagrangian description in
terms of
gauged WZNW theories \gwz\karabali\bcr.
Once the existence of a \sm description for a given  coset conformal theory  is
established
by considering a weak-coupling limit \bcr\wit\ one can  employ an `operator'
\dvv\Jack\bsfet\aat\at\
or `field-theoretic' \tsw\bs\aat\ approaches to  compute the exact form of the
\sm couplings.
While  in the bosonic case the leading-order (`semiclassical') solution is
modified by
$\a'$ corrections,  in the (1,1) supersymmetric (superstring) case the exact
solution  coincides with
the  semiclassical one \Jack\bsfet\aat.

The (bosonic) solutions corresponding to the coset c.f.t.'s provide the first
known examples of
exact string  backgrounds
 that  depend non-trivially  on $\a'$.  It is quite remarkable that
the properties   of the leading order and exact  forms of the
solutions may be quite different.  For  example,
  the causal structure of the exact  $SL(2,R)/U(1)$ solution  \dvv\ is
different  \bsfet\tv\py\  from that of the leading
order  one \wit ; in particular,  it was  claimed  \py\ that  the exact
Minkowski background   can be
represented in such a way  that it does not have the `black hole' singularity.
One  may  try to
interpret  this example as an indication    that the string $\a'$ corrections
may  remove  the
singularities of the Einstein theory solutions. One  should,  however, be
rather  cautious  about
the physical relevance of the $\a'$ corrections  in the present context  given
that  they are absent
in the superstring case:  the  exact solution in the   (1,1) supersymmetric
model   still has the
leading order `black hole' form \Jack.

This  raises the important question  of whether the $\a'$ corrections are
actually present in the
more `realistic'  heterotic string case \hh.
 In this paper
we are  going to  address this question by  discussing   a generalisation of
the  `operator' and
`field-theoretic' approaches  to
 derivation of the exact form of the coset-type  solutions  of the heterotic
string   described
by (1,0) supersymmetric sigma models (for  previous discussions of the
heterotic string in this
context see \bsf\bsfet\bb). The leading-order form of  such  heterotic string
solutions  is the same
as in the bosonic and (1,1) supersymmetric cases.
 A suggestion of how  to  find the exact form of the solutions
by directly combining the supersymmetric left and bosonic right sectors in the
 conformal field
theory  stress tensor was previously made in \bsfet\ with the conclusion that
the semiclassical solution, as in the bosonic case,  should be  modified by
corrections of all
orders in  $\a'$ or $1/k$. As we shall explain below, the approach of \bsfet\
does not seem to be
consistent
 with the perturbation theory analysis (similar to the one carried out for  the
bosonic and
supersymmetric cases in \ts\Jack) as well as with the Lagrangian approach based
on a gauged WZNW
theory (which is anomalous in the chiral (1,0) supersymmetric case).

We shall find that a consistent solution  exists if one identifies   the
heterotic  (1,0) \sm with the
supersymmetric (1,1)  one by  introducing the target space gauge field equal to
 the  Lorentz
connection.  This is equivalent to a particular example of   constructing a
left-right
symmetric solution of  the heterotic string  from a solution of the superstring
theory \chsw\hul\gep.
 As in the superstring case the
 leading-order solution is  then not modified by $\a'$ corrections.
If, instead, one sets the gauge
field to zero the leading-order solution receives  perturbative corrections  to
all orders in
$\a'$. It is not clear, however,  how to  sum them to an exact form since in
this case
the identification of the corresponding conformal field theory remains an open
problem.

 As in the  (1,1)  supersymmetric case the
simplest problem is to try to find the analog of the $D=2$ bosonic
$SL(2,R)/U(1)$ solution \bcr\wit.
One can  solve the corresponding `$\b$-function' equations  order by order in
$\a'$
determining the corrections to the leading-order $D=2$  background of \efr.
 In Sect.2 we shall  repeat the   analysis  of
 the  $D=2$ solution of the 3-loop `$\b$-function' equations  (carried out in
the bosonic case in
\ts\  and in the supersymmetric case  in \Jack) in the heterotic string case.
The  solution we shall find will  be different from the background which
corresponds to the
`heterotic' coset c.f.t. construction suggested   in \bsfet.

In Sect.3 we shall address the problem of constructing  a consistent  heterotic
string solution
with a clear  c.f.t. interpretation: its non-trivial  part will be described
by the (1,1) superconformal $G/H$ coset theory.   We shall use the Lagrangian
approach
presenting  a `heterotic' analog of the gauged WZNW model. The direct (1,0)
supersymmetric
truncation of the (1,1) supersymmetric gauged  WZNW theory  \bsf\ is  found to
be anomalous and does
not   correspond to a  consistent  heterotic string vacuum. The world sheet
anomaly can be cancelled
out by introducing the  coupling of the `internal' (1,0) spinor superfields to
the  target space gauge
field which is equal to the   coset model related Lorentz connection with
torsion.
 We shall first  discuss in some detail the  quantisation of the (1,1)
supersymmetric gauged WZNW model
(using the manifestly supersymmetric approach of \aat)  and  then demonstrate
that  its  (1,0)
supersymmetric truncation   is  not gauge invariant  at the quantum level.

Some duality properties of the exact $SL(2,R)/U(1)$ heterotic solution
will be studied in Sect.4.
We will  demonstrate  that not only the axial-vector duality relating the axial
coset model $SL(2)/U(1)_a$
to the vector  one  $SL(2)/U(1)_v$ is an exact symmetry of c.f.t and string
theory (as was shown in \GK), but here it also relates one {\it exact}
heterotic string solution to another {\it exact}
solution when acting on the curved
background matrix by a fractional linear transformation as in \GMR\GR.
Moreover, acting on the background matrix  as in \GMR\Ve\GR\ the full $O(1,2)$
rotations generate
exact backgrounds from exact backgrounds.
\newsec { $D=2$ perturbative  solution   and comparison with `heterotic' coset
conformal field theory construction}
 Let us  first review the structure of the heterotic string \sm
and the corresponding $\b$-functions and effective action.
In terms of (1,0) superfields we have \call\hw
$$ I = {1 \over \pi \a' } \int d^2 z d\t  \ [ G_{\m \n } (X)  + B_{\m \n } (X)]
DX^\m \bd  X^\n
+   \int d^2 z d\t  \Psi^I ( \delta^I_J D +   {\cal A}^I_{J\m} DX^\m ) \Psi^J \
, \ \
 \eq{2.1} $$
where
$$ X^\m= x^\m + { \t }{ \psi}^\m_+  \ , \ \ \  \ \Psi^I= \psi^I_- + { \t }{
f}^I  \ \ ,  \ \ \ D=
{\del\ov \del \t}- \t {\del\ov \del z}\ , \ \ \  \bd = {\del\ov \del \bar z} \
.  \eq{2.2} $$
The `external' world sheet (or, equivalently,  target space) chiral anomaly
can be cancelled by
making $B_{\m\n}$ transform under  the target space gauge  and Lorentz
transformations \hw.  This
implies that  the corresponding conformal anomaly coefficients and the
effective action which
generates them  should   depend on the antisymmetric tensor field
strength  in  gauge-invariant combination  with the  Yang-Mills  and Lorentz
Chern-Simons terms
$${\hat H}_{\m\n\l}   = H_{\m\n\l}  + {3\ov 4}  \a'\Tr
[ ( {\cal A}_{[ \m } F_{\n\l ]}  - {1\ov 3 } {\cal A}_{[\m} {\cal A}_\n
 {\cal A}_{\l]} ) -   ( \ho_{[ \m } {\hat R}_{\n\l ]}  -
 {1\ov 3 } \ho_{[\m }\ho_\n \ho_{\l]} )] \ \ , $$ $$
\ \ \ H_{\m\n\l} = 3 \del_{[ \m } B_{\n\l ]} \ . $$   If the gauge field
${{\cal A}}_\m$ is  equal to
the connection with torsion $\ho_\m$  (assuming that the  Lorentz group  can be
 embedded into the
gauge group \chsw) $$ {{\cal A}}_\m = \ho_\m \ \ ,\ \ \ \ \ \ho^i_{j\m }=
\o^i_{j\m}
   + \ha  H^i_{\ j \m } \ , \ \ \ \ \  H^i_{\ j \m } = H_{\n\l\m } E^{i \n}
E_{j}^{\l}   \ \ ,
\eq{2.3}  $$
 then
the \sm (2.1)   becomes (1,1) supersymmetric \hw\hul. The corresponding
`$\b$-functions' and effective
action  then become identical to those of  superstring theory.
Thus given a  superstring solution  one can always construct a particular
(1,1) supersymmetric solution of the heterotic  string theory with the
non-trivial  gauge field
background (2.3)\foot { It is always possible to embed the Lorentz group into
the internal gauge
group since the latter is big enough. There is also an important  issue of
modular invariance of
the solutions thus obtained. This question can presumably be treated along the
lines of ref. \gep. }.

The  first several leading order terms  in the heterotic string effective
action are given by
\hh\cal\het\met\gro\
$$  S = \int d^{D}x \sqrt {\mathstrut G} \ {\rm e}^{- 2
\phi} \lbrace c -  \alpha^{\prime} [R + 4
(\partial_{\mu} \phi)^{2}
- {1 \over 12} {\hat H}^{2}_{\lambda \mu \nu}]
$$ $$  + {1 \over 8} \a'^2 (  R_{\m\n\l\k} R ^{\m\n\k\l} -   \Tr
F_{\m\n}F^{\m\n} )
+ O(\a'^4)
\rbrace \ , \eq{2.4} $$
where $c$ vanishes at the critical dimension (but will be kept arbitrary in the
context of a general
(1,0) supersymmetric \sm (2.1))\foot
 {For discussions of `non-critical' heterotic superstrings see \aps\gny.  One
can keep $c$
arbitrary by considering (2.4) as a generating functional for the  conformal
anomaly coefficients of
a general (1,0) supersymmetric \sm (with the Lorentz anomaly being cancelled
by adding some
extra free chiral degrees of freedom).}.  For the particular background (2.3)
the action (2.4)
coincides (after a proper field redefinition) with the superstring effective
action. The important
point is that in contrast to the bosonic case there is no explicit $O(\a'^3)$
term in the heterotic
string effective action (2.4) \met.  As a result, there exists a  choice of a
scheme (or a field
redefinition) in   which the corresponding term is absent also  in the
conformal anomaly
coefficient for $G_{\m\n}$ \foot { As was  found  in \foa (see also \ell\ket)
the $O(\a'^3)$ term in
the $\b$-function  computed in the dimensional regularisation/minimal
subtraction scheme is
proportional  to the lower order terms in $\b$ and  corresponds to the  term
$\a'^3 R_{\m\n} (
R^\m_{\k\l \r} R^{ \n\k\r \l} - \Tr F^\m_\l F^{\n\l})$ in the effective action.
The latter   can be
redefined away (see also below). }    $$ {{\bar \b}^G}_{\mu \nu }= \a' (R_{\mu
\nu} - \fourth {\hat
H}_{\m\k\l} {\hat H}_\n^{\ \k\l}  +
 2  D_{\mu} D_{\nu} \phi   )
  + {1\over 4} {\a'}^2 (R_{\mu \a \b \g} R_{\nu}^{\ \a \b \g}  -  \Tr F_{\m \l}
F^\l_{\ \n }  ) +
O(\a'^4)  \ . \ \ \eq{2.5} $$
Let us follow \efr\ts\ and look for  a Euclidean  static  solution of (2.5)
with  $D=2$  target
space  (times some flat or toroidal  dimensions)
$$ds^2 =   dx^2 + a^{2}(x)\ d{\theta}^2  \ \  , \ \ \ \ \ \
a = {\rm e}^{\l (x)}\ ,\  \ \ \ \gp = \gp (x) \  . \eqno {(2.6)}$$
In two dimensions ${\hat H}_{\m\n\l}=H_{\m\n\l}=0$. We shall also  assume
first that the gauge field
is trivial,  $ {\cal A}_\m =0$.  Then the leading order solution of
$$ {{\bar \b}^G}_{\mu \nu }= \a' (R_{\mu \nu} +
 2  D_{\mu} D_{\nu} \phi   )
  + {1\over 4} {\a'}^2 R_{\mu \a \b \g} R_{\nu}^{\ga \gb \gg}  +
O(\a'^4) =0 \  \ \ \eq{2.7} $$
 is the same as in the bosonic and (1,1)
supersymmetric cases \efr
  $$ \ \gl = \gl_0 + {\ln \th}\  bx \ , \ \ \ \  \gp = \gp_0 - {\ln \cosh \
}  bx \ , \  \ \  {\ga'} b^2 = -c  \ \ .\  \eqno
{(2.8)}$$
The difference between (2.7) and the corresponding equation in the bosonic case
is in the coefficient (1/4 instead of 1/2) of the $RR$-term and in the absence
of the $O(\a'^3)$
terms.
Repeating the computation  of \ts\  for (2.7)  one finds the following
perturbative solution
$$ \gl =\gl_0 + {\ln \th}\ bx + {\ga'} b^2 (a_1  + a_2 {\ga'} b^2 ) {\th^2} bx
+
a_3{
\ga'^2} b^4 {\th^4} bx + O(\ga'^3) \ \ , \eqno {(2.9)}$$
$$ \gp =\gp_0 - {\ln \cosh \ } bx + { \ga'}b^2 (a_4 +  a_5{\ga'} b^2){ \th^2}
bx +
{a_6 \ga'^2} b^4{ \th^4 } bx + O(\ga'^3) \ \ . \eqno {(2.10)} $$
While  in the bosonic case \ts
$$ a_1 = 1\ , \ \ a_2 = -3\ , \ \ a_3 = 2\ , \ \ a_4 = \ha\ , \ \ a_5 = -\ha \
, \ \ a_6 = \ha\ , \ \
  $$
 in the heterotic case
$$ a_1 = \ha \ , \ \ a_2 = -23/18\ , \ \ a_3 = 41/72\ , \ \ a_4 =  \fourth \ ,
\ \ a_5 = -5/12 \ , \ \
a_6 = 25/144\ . \ \
 \eq{2.11} $$
 As  was shown in \ts\
the coefficients $a_2$ and $a_3$  in (2.9) can be changed  by a local field
redefinition to
$$  a_2' = a_2 +  q\ , \ \ \ a_3' = a_3 -   q\ \ ,   $$
where $q$ is arbitrary.
Let us now try to use this freedom  to  represent the perturbative solution
(2.9)  as an expansion of
the exact form (similar  to the one found in the bosonic case \dvv)   $$ \gl=
\l_0  +
{\ln \th}\ bx   -\ {\ha} {\ln}(1
 + \g  {\th}^2 bx)   \ ,   \ \eqno {(2.12)}$$
$$ \g = p_1 \a' b^2 + p_2 \a'^2b^4 + O(\a'^3) \ . $$
Then
$$\l =\l_0  + {\ln \th}\ bx  -\ha   { \ga'}b^2 ( p_1  +  p_2 {\ga'} b^2){
\th^2} bx +
\fourth p_1^2{ \ga'^2} b^4{ \th^4 } bx + O(\ga'^3) \ \ , \eqno {(2.13)} $$
$$ a_3  - q = \fourth p_1^2 \ \ , \ \ \   p_2 = -2 (a_2 + q) \ . \eq{2.14} $$
In the bosonic case  one finds  \dvv\ts
$$  p_1 = -2 \ , \ \
\  \ p_2 = 4 \ , \ \ \ \ q=1 \  ,   $$
which is consistent with  \dvv\ts
$$ \g = - {2{\ga'}b^2 \over 1 +2 {\ga'} b^2} \ \ .   $$
In the heterotic case  we get
$$  p_1 = -1 \ ,
\ \ \ \ \ p_2 = 23/12 \ ,  \ \ \ \ q = 23/72 \ \ .  \eq{2.15}  $$
For completeness let us present the  solution  that can be found by  starting
directly with the $\b$-function  of the heterotic string \sm computed in \foa\
without doing any
coupling redefinition.  For  $F_{\m\n} =0$ the  heterotic 3-loop $\b$-function
is given by \foa
$$ {{\bar \b}^G}_{\mu \nu }= \a' ( R_{\mu \nu} +
 2  D_{\mu} D_{\nu} \phi   )
  + {1\over 4} {\a'}^2 R_{\mu \a \b \g} R_{\nu}^{\ga \gb \gg}   $$ $$
+ {1\over 16} {\ga'}^3 [ \ha  D^2 (R_{\mu \a \b \g} R_{\nu}^{\ga \gb \gg})
 + 3 R_{\m \l \n \k } R^{\l \r\s\tau } R^\k_{\r\s\tau}  $$ $$
+ 2 R_{\m\k\l\r}(D^\k D^\l R^\r_\n - D_\n D^\l R^{\k\r})
- R_\l^\s R_\m^{\ \l\r\k } R_{\n \s\r\k} -
R_{\m \l}  R^{\l \s\r\k } R_{\n \s\r\k}]
+ O(\a'^4) \ . \ \ \eq{2.16} $$
Restricted to $D=2$,  eq.(2.16) reads
$$ {{\bar \b}^G}_{\mu \nu }=   \ha G_{\mu \nu} [ \a' R + \fourth \a'^2 R^2    +
{1\ov 32 }\a'^3 ( \del_\l R \del^\l R + s R^3 )  + O(\a'^4)]  +
 2  D_{\mu} D_{\nu} \phi   \ \ ,  \eq{2.17}  $$
where $s$ parametrises certain    ambiguity  in the results of \foa (one finds
$s= \ha $  and $s=5/2$
if one starts  respectively with  eq.(15) (or (2.16)) and eq.(14) of the first
paper in \foa).
In contrast to the bosonic case,  (2.17) does not contain the term $\del_\m R
\del_\n R $.
As in the bosonic case,  there may also be in principle the `diffeomorphism
term' $ D_\m D_\n  R^2 $
(which is absent in \foa ). Such term, however, is not important since it can
be absorbed into a
redefinition of the dilaton.
  Starting with (2.17)  we get
$$ a_1 = \ha \ , \ \ a_2 = (4s-11)/9 \ , \ \ a_3 = (-8s + 49)/72
 \ , \ \  $$ $$  a_4 =  1/4 \ , \ \ a_5 =
(s-2)/6  \ , \ \ a_6 = (-4s + 29)/144 \ ,
  $$
and hence
$$  p_1 = -1 \ , \ \
 \ \ \ p_2 =   (-8s + 19)/12 \ , \ \  \ \  q = (-8s + 31)/72  \ \ . \eq{2.15'}
$$
These values of $p_1$ and $p_2$ are in agreement with (2.15) if $s=-1/2$
(the values of $q$ are different because of an extra field redefinition
involved).


Let us now compare  (2.9),(2.10) with what one  would expect to find  for the
exact
background from  the operator approach assuming that there is a  corresponding
coset-type  conformal
field theory behind this solution (as   was the case  in the bosonic and (1,1)
supersymmetric
theories).   Let us first recall the situation in the
  bosonic  and supersymmetric  cases \dvv\Jack\bsfet.  The operators $L_0$
and   $\H= L_0 + {\bar L}_0  $ of  the $G/H$ coset conformal theory restricted
to the  zero mode
scalar sector  are  given by
 $$  L_0 = {1\ov k+ \ha c_G } J_G^2 - {1\ov k+ \ha c_H } J_H^2 \ \ , \ \ \ \  \
\
 \H = {2\ov k+ \ha c_G }[  J_{G/H}^2  - \g  J_H^2 ] \ \ ,  \eq{2.18} $$
$$  J_{G/H}^2 = J_{G}^2-J_{H}^2\ \ , \ \ \ \ \ \ \g =   {c_G - c_H \ov 2k+ c_H}
\ \  .  \ \
 $$
In the $SL(2,R)/U(1)$ case $\g  $  appears as  the coefficient  in the exact
metric (2.12) with
$\a'$ being  proportional  to  the overall coefficient in $\H$  (we  shall set
$b=1$)
 $$\ \a' = {1\ov k+
\ha c_G}  \ \ , \ \ \ \  \g = - {2\a'  \ov 1 +  2\a' } = - {2\ov k}  \   \ \ \
  . \eq{2.19} $$
 In the superstring case  $k$ is not shifted at all, i.e.
$$L_0 = {1\ov k }
J_G^2 - {1\ov k} J_H^2 \ , \ \ \ \H =
 {2\ov k}  J_{G/H}^2   \  ,  \ \ \ \ \g =0 \ , \ \ \ \ \a' = {1\ov k} \ .
\eq{2.20} $$
Since $\g=0$ the exact  metric is equal  to the semiclassical one.
In  both  the bosonic and  supersymmetric cases the left and right  parts of
the stress tensor  are
the same and the constraint  $(L_0 -{ \bar L}_0) T  = 0 $ (in the zero-mode
sector) is trivially
satisfied once the $H$-invariance constraint  $(J_H - {\bar J}_H) T  =0 $ is
imposed.

 The meaning  of this
constraint becomes less obvious when one tries to generalise this construction
to the heterotic case.
A  guess is to  take  the  operator $\H= L_0 + {\bar L}_0$  as  the combination
of the
supersymmetric ${ L}_0$  and bosonic ${\bar L}_0$  operators  assuming  still
that  $(J_H
- {\bar J}_H) T  =0 $  \bsfet.   Then (we omit possible `internal' part of
$\H$)\foot { Our expression
for the `heterotic' $\H$ is different from the one suggested in  \bsfet\ where
the shifted  $k + \ha
c_G$  was (inconsistently) used instead of the unshifted $k$ in the
supersymmetric
 part of the total operator.  Both the above expression  and the expression in
\bsfet\
$$ \H = {2\ov k + \ha c_G} [  J_{G/H}^2  - \g  J_H^2 ] \ , \ \ \a'=
{1\ov k + \ha c_G}= {1\ov k -2} \  , \ \ $$ $$  \g  =    {c_G - c_H \ov 4 k+
2c_H} =  -  {  1\ov k
} =   - {{\ga'} \over 1 +2 {\ga'} }= - \a' + 2 \a'^2 + ...  $$
 are in any case in  disagreement
with the perturbation theory  result for the corrections to the leading-order
solution.}
$$ \H=
L_0^{(susy)} + {\bar L}_0^{(bose)} = [{1\ov k } { J}_G^2 - {1\ov k} { J}_H^2] +
[{1\ov k+ \ha c_G } {\bar J}_G^2 - {1\ov k+ \ha c_H }{\bar  J}_H^2]
 $$
$$= ({1\ov k}  + {1\ov k + \ha c_G}) [  J_{G/H}^2  - \g  J_H^2 ] \ \ ,
\eq{2.21} $$
$$ \a' = {k + \fourth c_G \ov k(k+ \ha c_G) } \ \ , \ \ \
\g =   {(c_G - c_H)k \ov (2k+  c_H) (2k + \ha c_G ) } \ \ . \ \
 \eq{2.22} $$
In the $SL(2,R)/U(1)$ case
$$ \a' = {(k-1) \ov k (k-2) } \ \ , \ \ \ \
\g = - {1\ov k-1}
 =  - { 2\a' \ov 1 + \sqrt {1 + 4\a'^2} } =
-  \a'  +   \a'^3 + O(\a'^4) \ \ . \eq{2.23} $$
Let us compare this with the expression for $\g$ in the bosonic case (2.19),
$$ \g = -2\a' + 4 \a'^2 +
O(\a'^3)\ . $$
In contrast to the bosonic case, the  leading term in the heterotic $\g$
is  factor of 2 smaller and $\g$  does not contain the $\a'^2$ correction.
This   is reminiscent of similar differences in the structure
of the bosonic and heterotic effective actions. However, though the leading
term in $\g$ is indeed
in agreement with the perturbative solution  (2.13),(2.15) found above,   the
absence of the
$\a'^2$-term in $\g$ is in contradiction with the perturbative solution. As it
is clear  also from
(2.15$'$), $p_2$ does not vanish for  the values of $s$ quoted above. This
suggests that the naive `heterotic' construction of the  coset conformal field
theory  does not
actually work.

In  the above discussion of the  perturbative solution we have assumed that the
gauge field background
is trivial. There  should  exist also the   heterotic  string solutions   with
non-vanishing gauge
field backgrounds.   One  obvious possibility is to set the gauge field equal
to the Lorentz
connection returning effectively to the (1,1) supersymmetric  superstring case.
 Such solution will
have a simple c.f.t. counterpart with  the $\H$-operator  of the conformal
theory
being equal to the  one in the  supersymmetric  case  (2.20).   Therefore,  the
{\it exact}
heterotic string solution will be given by the semiclassical background
(2.6),(2.8) and the abelian
gauge field 1-form
 $$ {\cal A}_r =0\ , \ \ \ \  {\cal A}_\t = \o^r_\t = -  {b\ \e {\l_0}  \ov
{\rm cosh}^2
bx } \ \ . \eq{2.24} $$
 Let us note that this solution is different from the perturbative  charged
black hole solution of the $D=2$ heterotic string theory found in
\gny.  The truncation of the effective action  (2.4) used in \gny\
corresponds   to  treating  the gauge field term  on an equal footing with  the
$\a'$ terms in (2.4) while dropping out  all other terms of higher orders in
$\a'$.
At the same time, the  (1,1)
supersymmetric  solution based on identifying   the Lorentz connection  with
the gauge field  (2.3)
is exact because of  the   cancellation  of all higher order  gravitational
corrections  against the
 gauge field  dependent ones. While  higher order terms  in the  superstring
effective action or
in the $\b$-functions  are  known to be  present  for a  general  background
\gr\ they actually
disappear (in a particular scheme) for the backgrounds  corresponding to coset
the c.f.t.'s
\Jack\bsfet\aat.

\newsec { Exact solutions  based on `heterotic' gauged  WZNW theory}
Below we shall consider the construction  of the left-right symmetric
coset-type heterotic
solutions using the field-theoretic or Lagrangian approach. The  basic element
will be
the   (1,0) supersymmetric analog of the gauged WZNW model.
Let us
start with a review of  the  manifestly supersymmetric approach to
 the quantisation of the (1,1) supersymmetric  gauged WZNW model  which was
suggested  in \aat.
The   superfield form of the (1,1) supersymmetric  gauged WZNW action  is
$$ \hat I (\hg, \hat A) = \hat I(\hg )  +{1\over \pi }
 \int d^2 z d^2 \t  \Tr \bigl( - \hat A\,\bar D \hg \hg\inv +
 {\hat {\bar A}} \,\hg\inv D \hg + \hg\inv \hat A \hg {\hat {\bar A}}
 - \hat A {\hat {\bar A }} \bigr)\  \ ,  \eq{3.1} $$
where  the  gauge superfields $\hat A , \hat { \A}$   take values in the
algebra of  the subgroup
$H$. The supersymmetric  version  $\hat I(\hg )$ of the ungauged WZNW action
$I(g)$
is obtained by
 replacing  the  group field $g$   by the
 corresponding superfield and $z^a\ra (z^a,
\t , \bar \t ) $,
$ \ \del \ra D$, etc.  \div
$$\hat g= \exp (T_A X^A) \ , \  \ \ \ X^A= x^A + { \t }{ \psi}^A_+ + {\bar \t
}{ \psi}^A_-  + \bar \t \t f^A \ , \ \ \ D= {\del\ov \del \t}- \t {\del\ov
\del z}\ , \eq{3.2} $$
  $$ \hat S=
{k }  \hat I(\hg) \  ,  \ \  \   \  \hat I(\hg)  \equiv  {1\over 2\pi }
\int d^2 z d^2\t   \Tr \{ D \hg^{-1}
\bar D \hg   - {i}   \int dt [\hg^{-1} D\hg , \hg^{-1} \del_t\hg ] \hg^{-1}
{\bar D} \hg  \ \} \
  \ . \eq{3.3} $$
Parametrising   $\hat A$ and $\hat \A$ in terms of the superfields
$ \hat h$ and $\hat \bh$  from $H$
$$ \hat A = \hat h D {\hat h}\inv \
\ , \ \ \ \hat \A = \hat{ \bh }\bar D  {\hat \bh }\inv  \ \ , \eq{3.4} $$
we get
$$ \hat I (\hg, \hat A)= \hat I(\tilde \hg) - \hat I(\tilde {\hat h}) \ ,
\eq{3.5} $$
$$
\tilde \hg\equiv {\hat h}\inv \hg \hat{\bh}\ , \ \ \ \
\tilde {\hat h} \equiv {\hat h}\inv \hat \bh \ . \eq{3.6} $$
The  action and the variables $\tilde \hg$ and $\tilde {\hat h}$ are invariant
under the superfield
gauge transformations parametrised by  a superfield $\hat u$ with values in
$H$.
Similarly  to what happens in the bosonic case \gwz\karabali\ the quantisation
of the theory can
thus  be reduced to that of the
two
ungauged supersymmetric WZNW theories corresponding to the group $G$ and the
subgroup  $H$
$$ Z = \int [d\hg] [d\hat A][d\hat {\A}] \ \exp \{ - k \hat I(\hg , \hat A) \}
  $$
$$ =  \int [d\tilde {\hg }] [d\tilde {\hat h}]  \ {\cal J}  \ \exp \{  - k I
(\tilde {\hat g} ) +
   k  I (\tilde {\hat h})    \} \ \ . \eq{3.7} $$
Here  the functional measure includes a gauge fixing factor and $\cal J $
stands for  the product of
Jacobians of the change of superfield variables
from  $\hat A$ to  $\hat h$  and from $ \hat {\A} $ to $ \hat {\bh} $. While in
the bosonic case
the  corresponding product  (regularised in the left-right symmetric way) is
non-trivial and leads to
the shift  of the coefficient  $-k \ra  -k - c_H$
of the $H$-term in the action \karabali ($c_H$ is the value of the second
Casimir of $H$),  in the
(1,1) superfield  case   each of the Jacobians is   proportional
to  a field-independent factor.  In fact,  as in the bosonic case,
the Jacobian of the change $\hat A \ra \hat h $
can be expressed in terms of the  path integral
$$ \int [dU] [dV] \exp \{ - \int d^2 z d^2 \t  U ( D + [ \hat A , \ ])V \ \}  \
\ , \eq{3.8} $$
where  $U$ and $V$ now are  the (1,1) superfields of opposite statistics.
Rewriting  the  action  in
(3.8) in  component fields   and
integrating them out  it is easy to see that this Jacobian is $\hat
A$-independent:  the
non-trivial contributions of the bosonic and  fermionic
determinants   are equal and cancel each other.
The theory  can thus be  represented  as a `product' of the two  (1,1)
supersymmetric
ungauged WZNW theories for
the  groups  $G$ and $H$   with   the levels $k$ and $-k$.
As discussed in \aat\ this implies that there is no finite renormalisation of
$k$ in the
corresponding quantum effective action  which is  thus equal (up to unimportant
field redefinition)
to the classical action.  Therefore, the  expression for the exact background
coincides with the
leading order one  without any $\a'$ correction.

It is instructive to reformulate the above discussion in terms of the component
fields.
Let us recall that the ungauged supersymmetric WZNW theory $kI(\hat g) $ can be
 represented as the
sum  \sus\ of the   bosonic WZNW action with shifted level $k-\ha c_G$
\suss\red\schn\ and
free fermions.
 The shift of $k$ originates from the anomalous  contribution of the  chiral
rotation of the (right)
fermions neeeded to decouple them from the bosons.  Then the component form of
(3.7) is
$$ Z =   \int [d\tg][d\tth ][d\psi_G]
[d\psi_H]    \  \exp [ - (k- \ha c_G)I(\tg)   + ( k +  \ha c_H)I(\tth)  $$
$$  - k I_0(\psi_G)  +   k I_0
(\psi_H) ] \ \ ,  \ \eq{3.9} $$
where $
\tilde g\equiv { h}\inv g {\bh}\ ,  \ \
\tilde { h} \equiv { h}\inv  \bh \ .  $
 Up to the free-theory  factors,
 the
resulting theory  can be represented as  the `ratio'  $G_{k-\ha c_G}/H_{k - \ha
c_H}$ of the
bosonic WZNW theories for
the groups $G$ and $H$   with levels  $k_G= k-\ha c_G$ and  $k_H= k - \ha c_H$
(we separate
 the shift $c_H$  corresponding to the  bosonic change of variable from $k_H$).
This  conclusion is  in
agreement  with   the   conformal algebra approach \gko\fuch\ks (note that in
terms
of the shifted level
${\hat k}=k-\ha c_G$    we get the  $G_{{\hat
k}}/H_{{\hat k} + \ha (c_G -  c_H)}$   theory).

Let us now compare the  above approach with the one in which one starts
directly with
the component
formulation of the gauged  supersymmetric WZNW  theory where the fermions are
coupled only to $A,
\A$  (see e.g. \schn\swz\bsf).   If one integrates over the fermionic
components of  the gauge
superfields $\hat A$ and $\hat \A$ in  (3.1) one  still has the fermionic gauge
invariance in
the resulting action. The latter can be fixed by  setting the $H$-components of
the
fermionic part of $\hat g$ to zero as a gauge.  Making a (chiral) rotation to
decouple the fermions
from $g$  (and ignoring the corresponding shift of $k$) one then finishes with
the following classical
action $$  I (g,A,\psi_{G/H}) =  k I(g,A) + k I_0(\psi_{G/H},A) \ \ ,
 \eq{3.10}
$$
$$ I_0(\psi_{G/H},A) = \int d^2z\  \psi_+ (\bd + [\A, \ ] )\psi_+
+   \int d^2z \ \psi_- (\del + [A, \ ] )\psi_- \ \  . \eq{3.11} $$
Here $I(g,A)$ is the bosonic WZNW action  and $I_0(\psi_{G/H},A)$ is the action
of
the Majorana fermions $\psi^i_{G/H} = (\psi^i_+, \psi^i_-)$
taking values in the orthogonal complement of the algebra of $H$ in the
algebra of $G$.
Once the gauge field is integrated out and the  bosonic gauge  invariance is
fixed by
restricting $g$ to $G/H$ one gets the equal number of bosonic and fermionic
degrees of freedom.
If one uses the   action (3.10)   as a starting point for quantisation of the
gauged
supersymmetric WZNW  theory  one finds
$$ Z= \int [dg][dA ][d\A] [d\psi] \ \exp \{- k I(g,A) - I_0(\psi,A) \} \ ,
\eq{3.12}
$$
 or after integrating over the fermions   and changing the
variables from $A,\A$ to $h,\bh$
$$ Z =   \int [d\tg][d\tth ]   \  {\exp \{ -  k I(\tg)   +  [ k  +
\ha (c_G - c_H) +   c_H] I(\tth)
\} } \ \ \  .  \ \eq{3.13} $$
The fermionic path integral  (regularised in the left-right symmetric way) gave
 the following
contribution \polwig\gwzw\oal
$$  [\det (\del + [A, \ ] ) \det (\bd + [ \A , \ ] )]^{1/2}
= \exp{ [\ha  c_K I(h\inv \bh ) ]} \det \del \det \bd \ , \ \ \ \ c_K = c_G -
c_H \ .\eq{3.14} $$
 The path integral  (3.13)
 becomes equivalent to (3.9)  if $k$ in (3.10),(3.13)  is replaced by $\hat k
= k - \ha c_G$.
As in the case of the ungauged supersymmetric WZNW theory
the  approach based on the `decoupled'
  component action (3.10) loses  the shift of $k$ and this is inconsistent
with  manifestly
supersymmetric perturbation theory  \aat.

Let us now turn to the heterotic case.
The corresponding (1,0) supersymmetric actions can be obtained by  a
truncation of the (1,1)
supersymmetric ones (i.e. by dropping the $\bar \t$ components of the
superfields and replacing
$\bar D \ra \bd, \  \int d \bar \t \bar D \ra \bd $, etc. ).
For example, let us start with the naive component action  (3.10) and drop out
the right part of
$\psi_{G/H}$ adding instead some  internal fermions $\psi^I_-$  which are not
coupled to $A, \A$
but may be coupled to a background target space gauge field ${\cal A}_\m$
(cf.(2.1)). The resulting
action (with ${\cal A}_\m=0$)
 $$  I (g,A,\psi_+, \psi_-) =  k I(g,A) + k ( \int d^2z \psi_+ (\bd + [\A, \
] )\psi_+  +   \int d^2z \psi_-^I \del \psi_-^I \ ) \ \   \eq{3.15}
$$
is equivalent to the one
  suggested  in \bsf\ as an action of  the heterotic string in a background
corresponding to  a  $G/H$  `coset'  conformal field theory.

The theory described by (3.15) is not, however,  well defined: since the
dynamical $2d$ gauge field
$A_m$ is coupled only to the left fermions  the resulting path integral is
anomalous\foot { It is
the {\it target
space}   anomaly that can be cancelled out  by modifying the transformation
rule of the
antisymmetric tensor field in  the standard (1,0) supersymmetric heterotic \sm\
(2.1) \hw.   Here, however, it is the $2d$ gauge symmetry corresponding to a
dynamical gauge field
$A_m$  that is anomalous. }.
 In fact,  if one replaces  the action  in (3.12)  by  (3.15),
changes  the bosonic variables and  integrates  over
 the fermions  $\psi_+, \psi_-$  (but  does not do  gauge fixing) one finds
(cf. (3.13))
$$ Z =   \int [d g][d h ][d\bh ]    \  {\exp \{ -  k I(h^{-1} g\bh )
+  ( k  +  c_H) I(h^{-1} \bh)   +
\ha (c_G - c_H) I(\bh)
\} } \ \ \  .  \ \eq{3.16} $$
In contrast to the result in the left-right symmetric  superstring  theory
(3.13) this
 path integral is not gauge invariant\foot { The anomaly is absent if
$c_G=c_H$.
Recall that this anomaly is  found
after using as a starting point  the  action (3.15) where the fermions  are in
$G/H$ and not in
$G$.}.
If one   still formally integrates over  $A$ and
$\A$ (or  $h$ and $\bh$)  one    obtains  an action for $g$ which cannot be
reduced to  a \sm action
with $G/H$  as a target space because of the lack of gauge invariance. Thus
the action (3.15) does
not   describe    a consistent   heterotic
string   background.

A similar   conclusion is reached    if one starts with the manifestly (1,1)
supersymmetric
action (3.1) and truncates the superfields to (1,0) superfields (adding also
some internal (1,0)
superfields $\Psi^I$ as in (2.1)). The resulting action is
$$ \hat I (\hg, \hat A, \Psi) = \hat I'(\hg )  +{1\over \pi }
 \int d^2 z d \t  \Tr \bigl( - \hat A\,\bar \del \hg \hg\inv +
 {\hat {\bar A}} \,\hg\inv D \hg + \hg\inv \hat A \hg {\hat {\bar A}}
 - \hat A {\hat {\bar A }} \bigr)\  + I_{int} \ ,
 \eq{3.17} $$
where $\hat I'$ denotes the (1,0) supersymmetric WZNW action (see e.g.
\hull\gat)
$$ \hat I'(\hg)  \equiv  {1\over 2\pi }
\int d^2 z d\t  \{ \Tr (D \hg^{-1}
\bd \hg )  - {i}   \int dt [\hg^{-1} D\hg , \hg^{-1} \del_t\hg ] \hg^{-1} \bd
\hg \  \} \ ,
\eq{3.18} $$
and
$$  \ \ \
I_{int} = \int d^2z d\t   \Psi^I ( \delta^I_J D +   {\cal A}^I_J\ ) \Psi^J \ ,
\ \ \ \
{\cal A}^I_J=  {\cal A}^I_{J\m} (X)  DX^\m  \eq{3.19} $$
$$\hat g= \exp (T_A X^A) \ , \  \ \ \ X^A= x^A + { \t }{ \psi}^A_+  \ , \ \ \ \
\ \Psi^I= \psi^I_- + { \t }{ f}^I  \ \ , $$ $$  \   \hat A =  \chi_+  + \t A  =
 \hat h D {\hat
h}\inv \ , \ \ \  \hat \A = \A + \t \chi_-  =\hat{ \bh }\bd  {\hat \bh }\inv
{}.
 $$
We have included the coupling of the internal fermionic superfield
to some background target space gauge field ${\cal A}^I_{J\m} (X)$  which may
depend on  some
combinations $X^\m $ of the
 (1,0) superfields  $\hg, \hat A, \hat \A$.  Changing the variables as in (3.7)
we get
 the following path integral
$$ Z = \int [d\hg] [d\hat h][d\hat {\bh}]   \ {\cal J}'  \ \exp \{  - k I
({\hat h}\inv \hg \hat{\bh} ) +
   k  I ({\hat h}\inv \hat \bh)    \} \ \ , \eq{3.20} $$
where ${\cal J}'$ is again the product of Jacobians of the two changes of
variables $ \hat A \ra \hat
h $
and $ \hat \A \ra \hat \bh$.  As in the  (1,1) case (3.8) these Jacobians   can
be represented   in
terms of  the path integrals  with the  (1,0) superfield actions ($U$ and $V$
have the same
statistics while $\bar U$ and $\bar V$  have  the opposite one)
$$ I=\int d^2 z d \t  U ( D + [ \hat A , \ ])V  \ \ , \ \ \
 \ \bar I = \int d^2 z d \t  \bar U ( \bd  + [ \hat \A , \ ])\bar V  \ \ .
\eq{3.21} $$
Re-writing these actions in terms of component fields it is easy to see that
the first Jacobian is essentially the same as in the bosonic case while the
second is still trivial
as in the (1,1) supersymmetric case. As a result,
$$ Z =  \int [d\hg] [d\hat h][d\hat {\bh}]   \   \ \exp \{  - k \hat I'
({\hat h}\inv \hg \hat{\bh} ) +
   k \hat  I' ({\hat h}\inv \hat \bh)    +  c_H  \hat I' ({\hat h}\inv )  \} \
\ . \eq{3.22} $$
There is  also an extra anomaly term originating from  non-invariance of the
path integral measure  under the (1,0) superfield rotations:
$$ [d\hg] [d\hat h][d\hat {\bh}] = [d({\hat h}\inv \hg \hat{\bh} )] [d ({\hat
h}\inv \hat \bh) ]
[d\hat h]
 \ \exp \{
  - \ha (c_G + c_H )  \hat  I' ({\hat h}\inv)     \} \ \ .
 $$
 As  in (3.16), the total anomaly
cancels out  if $c_G=c_H$.\foot { This may  imply a possibility  to   get  a
consistent topological theory  in the case when $H=G$. $G/G$ models can be
viewed as the infrared
(strong coupling) limit of gauged WZNW models allowing a kinetic energy term
for the gauge fields.
The case $G=U(1)$ can be followed for all values of the coupling.
Gauging an anomalous $U(1)$  \GG\ results, in the $A_0=0$ gauge, in having
massive
particles not obeying relativistic dispersion relations as well as having
massless chiral fermions. In the strong coupling limit the  massive sector
 decouples leaving just a topological theory and a
massless fermion. The remaining sector  does  not have enough structure  in
order to be able  to
inquire  about  unitarity and Lorentz invariance. Thus a chiral gauging
may be possible in general for $G/G$ models.
}

  The correspondence with (3.16) can be  traced  more directly as
follows.
 Let us  start with (3.22) and integrate over the fermionic parts of $ {\hat
h}\inv \hg$  and $\hat
h$ (or, equivalently, of $\hat g$ and $\hat h$),  treating $\bar h$ as a
bosonic background. Then we
will get an extra  term $  + \ha (c_G + c_H )   I ({\bar h})$ in the exponent
in (3.22).
Supersymmetrising this term and combining it with $ c_H  \hat I' ({\hat h}\inv
)$    (using the
superfield version of the Polyakov-Wiegmann relation \polwig\div\ and dropping
local terms) we get
(cf. (3.16))\foot
 { Let us note, however,  that, in principle,  the correspondence between the
quantum versions of the
 `reduced' component  theory (3.15) and the manifestly supersymmetric theory
(3.17)  may not
hold in the present anomalous situation. In fact, recall that  the classical
action (3.15)  can be
obtained  from (3.17)  by fixing  the fermionic part of the supergauge
symmetry. The latter,
however, is anomalous at the quantum level. }
$$  c_H  \hat I' ({\hat h}\inv ) + \ha (c_G + c_H )  \hat  I' ({\hat \bh})  =
  c_H  \hat I' ({\hat h}\inv \hat \bh ) + \ha (c_G - c_H )  \hat  I' ({\hat
\bh})  + ... \ \ . $$

We conclude  that the naive (1,0) supersymmetric version of the gauged WZNW
theory
with decoupled internal sector (i.e. a zero target space  gauge field) does not
describe a consistent
heterotic string background.  This is   related to what we have found  in
Sect.2:
the naive  `heterotic' construction of the  coset  c.f.t. operator (2.21)
which seems to  be
equivalent,  at the Lagrangian level, to the direct (1,0) supersymmetric
truncation of the gauged
WZNW theory  does not  correspond to a  perturbative solution  of the heterotic
string field equations.

In view of the discussion in Sect.2  a natural  possibility to obtain a $2d$
action describing
a consistent
 heterotic string solution is to introduce a target space gauge field
background
such that the resulting  $2d$ theory becomes effectively (1,1) supersymmetric
and  thus  anomaly free.  In this way any exact  superstring solution
corresponding to a  (1,1) supersymmetric coset c.f.t. or (1,1) supersymmetric
gauged WZNW theory  will
generate an exact solution of the heterotic string theory.  If one  takes the
internal fermions
$\psi^I_-$ in (3.15) to belong to the tangent space to $G/H$  and  couples them
to a  background
field  ${\cal A}_\m$ such that the  corresponding   $2d$ gauge field
 ${\cal A}= {\cal A}_\m \del x^\m$   is  equivalent  to  $A$ then (3.15)
becomes  identical  to the
(1,1) supersymmetric action  (3.10).  The role of the  coupling of the internal
fermions to  an
appropriate  ${\cal A}  $   is to ensure that   their  contribution
cancels the  total  $2d$ gauge anomaly.

To  formulate  the resulting model   in a more precise way  let us  first make
the following comment.
As in  the  case of the bosonic and (1,1) supersymmetric gauged WZNW models
one  may assume that  a
starting point  is a $2d$ theory with  some extra ($2d$ gauge) fields  while  a
 `sigma model'
(or `string in a background') interpretation is achieved only   after these
fields are integrated
out.  Adopting  this  point of view  one   may  consider the   (1,0)
supersymmetric model where the
internal fermions  are coupled   directly to the $2d$ dynamical field $A$ of
the (1,0) supersymmetric
gauged WZNW theory.  Such an action  will take  the form of  the    heterotic
string sigma model
action
 (in which   the internal fermions  are coupled to a particular $2d$ gauge
field
${\cal A}= {\cal A}_\m \del x^\m$, namely,  to the  projection of the target
space gauge field on
the world sheet)  only after the $2d$ dynamical gauge fields are integrated
out.
Alternatively,  one  may  start with  a more easily interpretable `heterotic'
action where   the
internal fermions  are coupled directly to
 a target space gauge field $\cal A$ (which is a function of $g$ only)
equivalent to
the `classical' value of $A$, i.e. $A_{class} (g)$   found by solving the
field equations of the
gauged WZNW model.   Since  the integral over $A, \A$ is gaussian (so that
the  integration  is  equivalent to replacing  $ A,  \A$ by their `classical'
values) the two
approaches  lead to  equivalent  actions  after the elimination of the $2d$
gauge field.

In the manifestly supersymmetric  formulation   (3.17) the second option
corresponds to  coupling
$\Psi^I$   to
 $${\cal A }= {\hat A}_{class}= {\hat A}(\hg) \ \  , $$
where ${\hat A}_{class}$   is the value one finds
by solving the classical   equations  for  the action  (3.17)
$$ \hat A^a_{class}  = -
M^{-1 ab } (\hg)  \Tr (T_b \hg\inv D \hg ) \ \ , \  \eq{3.23} $$
  $$  \ \ \hat A_{class}=
\hat A^a_{class}T_a\ ,   \ \ \ \ \ \ \ M_{ab } (g)  \equiv  \Tr (T_a g T_b
g\inv  - T_a T_b ) \
  $$   ($T_a$ are the generators of the subgroup $H$).
 Equivalently,  one may first  integrate over the
gauge field components in   (3.17)
obtaining the following (1,0) supersymmetric  \sm  (cf. (2.1))
$$ \hat I'' (\hg)  \equiv  {1\over 2\pi }
\int d^2 z d\t   \{ \Tr (D \hg^{-1}
\bd \hg )  - {i} \Tr   \int dt [\hg^{-1} D\hg , \hg^{-1} \del_t\hg ] \hg^{-1}
\bd \hg    $$
$$ +
   \Tr (T_a  \hg\inv  D \hg  )\  M^{-1ab} (\hat g )   \Tr
(T_b \bd \hg \hg\inv ) \
\} +  I_{int} \ \ . \
 \eq{3.24} $$
Since   the gauge field background is  chosen  to be  such that
the $2d$ gauge anomaly cancels out  we can  fix the gauge by restricting $\hat
g$ to $G/H$ (the   corresponding coordinate  (1,0) superfields  are $X^\m$).
Then  (3.24) reduces to
 $$ I = {1 \over \pi \a' } \int d^2 z d\t  \ [ G_{\m \n } (X)  + B_{\m \n }
(X)] DX^\m \bd
X^\n  $$ $$  +   \int d^2 z d\t  \Psi^I ( \delta^I_J D +   {\cal A}^I_{J\m}
DX^\m ) \Psi^J \ , \ \
 \eq{3.25} $$
$$ G_{\m\n}  = { G}_{0 \m\n }  -
 2 ( M^{-1})_{ab} E^a_{(\m }\E^b_{\n)} \ , \ \ \
   B_{\m\n} = B_{0 \m\n} -
 2( M^{-1})_{ab} E^a_{[\m }\E^b_{\n] } \ ,
\eq{3.26 } $$
where $G_0, B_0$ correspond to the   WZNW couplings  and $E^A_M$ and $\E^A_M$
are the left and right
vielbeins on  the group $G$  \bs\aat\foot{ There is also the dilaton background
  given by
$\p   = \p_0  - \ha  {\
\rm ln \ det \ } M \ $.}.  Identifying the
gauge field $\cal A$ with the connection with torsion corresponding to $
G_{\m\n} + B_{\m\n}$  (3.3)
one finally gets the (1,1) supersymmetric \sm $$ I = {1 \over \pi \a' } \int
d^2 z d^2\t  \ [ G_{\m \n
} (X)  + B_{\m \n } (X)] DX^\m {\bar D}  X^\n   \ ,  \eq{3.27} $$
where now $X^\m (\t, \bar \t) = X^\m (\t) + {\bar \t } \Psi^\m  + \t\bar \t
f^\m $ is a (1,1)
superfield. The    model (3.27)   can be obtained also directly   by
integrating over the  gauge
superfields  $\hat A, \hat \A$ in   the  (1,1) supersymmetric gauge WZNW
action (3.1).
Switching to the component notation   it is then easy to  read off
(from the  quadratic and quartic fermionic terms in the action)
 the corresponding connection
$\hat \o$  (or the gauge field, cf.(3.23)) and its curvature ${\hat R}$ (or the
gauge field
strength): $$ {\hat  \o}_\m= {\hat  \o}^a_\m T_a \ , \ \ \ \ {\hat  \o}^a_\m=
{\cal A}^a_\m = - M^{-1
a b}E_{ b\m}\ , \ \eq{3.28} $$
$$ {\hat R}^a_{\m\n} ={\hat F}^a_{\m\n} = - 2 \del_{[\m} M^{-1 a b} E_{b\n ]}
- M^{-1 a b} f_{b AB} E^A_{\m } E^B_\n  + f^a_{cd}  M^{-1 cb} M^{-1 de}
E_{b\m}E_{e\n}\ ,
\eq{3.29} $$
where $f_{ABC}$ are the structure constants of the group $G$.\foot { In the
general case of  the
model (3.27) the fermionic components $\psi_+$ and $\psi_-$ are coupled to the
generalised
connections $\o_-= \o - \ha H $ and $\o_+ = \o + \ha H$. Here ${\hat \o}=\o_+$
is the `right'
component of the  spin connection which together with ${\hat R}$ vanish in the
group space
(supersymmetric WZNW model) case  \sus\suss. }

At the level of conformal field theory this  exact `left-right symmetric'
solution is described by the
(1,1) supersymmetric $G/H$ coset c.f.t. The  non-trivial  part of the
corresponding   stress tensor
operator  is left-right symmetric and is the same as in the superstring theory
(2.20)  (cf.
(2.21)) $$  \H=
L_0^{(susy)} + {\bar L}_0^{(susy)}  = {1\ov k } [{ J}_G^2 -  { J}_H^2] +
{1\ov k} [{\bar J}_G^2 - {\bar  J}_H^2] = {2\ov k} J^2_{G/H}  \ . \eq{3.30}  $$
The  role of the gauge field background is effectively to convert the right
part of the stress tensor
operator into the supersymmetric one, making the  background metric
corresponding  to (3.30)
exactly  equal to the leading order one. This is consistent with what one
finds
 from the perturbation theory  analysis  under the identification (2.3).

 There are, however, other heterotic string solutions for
which  a  conformal field theory  interpretation  is not obvious. In Sect.2 we
have  presented  the
perturbative $D=2$ solution  (2.9)-(2.11)  with  the {\it   vanishing  }  gauge
field  (and
$B_{\m\n}$) background  and the  same leading-order metric and dilaton  as in
the bosonic and
superstring cases. Since it is not clear  which  c.f.t.    corresponds to
this solution we  are unable to  write down  its generalisation to all orders
in $\a'$.
A similar situation  is  known  in the case  of  $N=2$ supersymmetric
Calabi-Yau-type solutions of the
critical $D=10$  heterotic string: while  the `left-right symmetric' solutions
corresponding to the
embedding of the  Lorentz connection into the gauge group \chsw\gep\foot { Ref.
\gep\ contains a
string theory level description   of the general mechanism (essentially
equivalent to  the
embedding  of the Lorentz connection into the gauge group) by which every
superstring solution can be
converted into a  left-right symmetric solution of the heterotic theory.} are
described by  tensor
products of $N=2$ minimal models   \gep\  (with the  `left' and `right'
conformal theories being the
same)   other possible solutions (e.g. with vanishing gauge field or
non-trivial antisymmetric
tensor) which  should certainly exist as it is clear   from the
perturbation-theory  point of view
cannot be readily described  in c.f.t. terms\foot { Solutions with  (2.3) and
non-vanishing
antisymmetric tensor background  can
 probably be  described  as deformations of the  models of \gep.}.


\newsec {Exact duality in $SL(2)/U(1)$ heterotic background}

In this section we  shall discuss the target space duality relating the axial
$U(1)$ gauging of $SL(2)_k$  to the vector gauging  in  the  case of (1,1)
symmetric  heterotic
solution of the previous sections.
It can be shown (along the lines of
ref.\GK) that this
duality is an exact symmetry of c.f.t. and string theory corresponding
to a residual broken gauge symmetry.
In the bosonic string it relates  (either  at the \sm level
to the leading order in $\a'$ \mplt\  or at
the WZNW  or coset theory level \giv\dvv\kir)  the 2-d `black hole' solution
\bcr\efr\wit\ (the semi-infinite cigar) to its dual \giv\dvv\bsft (the
semi-infinite trumpet).

The extension of the results of \GK\ to the heterotic string is
immediate because  the gauge symmetry relating the axially gauged
abelian coset to the vectorially
gauged coset acts as a Weyl reflection either on the chiral current or the
antichiral current.
For  the heterotic string at the $SL(2)$ point,
a Weyl reflection of the antichiral current $\bar{J}$ in the
bosonic sector  will, therefore, generate a duality
symmetry along the line of $J\bar{J}$ deformation. At the
boundaries of the deformation line  this symmetry gives rise  to the axial
coset and vector coset
duality.

For the axially gauged  bosonic $SL(2)_k/U(1)_a$ euclidean model
 one gets (to leading order in $\alpha'$ or
in $1/k$) the  following line element and dilaton (cf.(2.6),(2.8))
$$
ds^2=k(dr^2+\tanh^2 r \ d\theta^2)\ , \ \ \  \phi=\phi_0- \ln \cosh r\
.\eq{4.1}
$$
In (4.1) $\theta$ is periodic with periodicity $2\pi$, and $\phi_0$ is a
constant.
Here  $r=x/{\sqrt{k}}=bx$ where $x$ is the coordinate in (2.6),(2.8) and
$e^{2\lambda_0}=k,\ \ b=e^{-\lambda_0}$.
In  what follows  we shall use the notation of \GMR\ and
\GR. We refer the reader to these references  for more details.

As discussed above,  the background  (4.1) corresponding to $SL(2)_k/U(1)_a$
with a
gauge field equal to the Lorentz connection  is an exact solution of the
heterotic string theory.
 The non-trivial part of the background  which transforms under the duality is
encoded in the $2\times 2$  matrix   with  $\theta,\ \theta_I$ components,
where $\theta_I$ is the internal coordinate corresponding to the embedding
of the $U(1)$ connection in the internal gauge group (see \GR\ for
more details).

The relevant block $E$ of the background matrix is therefore
$$
E=\left(\matrix {
G_{\theta\theta} & -2{\cal A}_{\theta}\cr  0 & 1
\cr}\right) =
\left(\matrix {
k\tanh^2 r & 2/\cosh^2 r \cr 0 & 1 \cr}
\right)\ .
\eq{4.2}
$$
Here ${\cal A}_{\theta}=\omega_{\theta}^r =-1/\cosh^2 r$.
We did not correct $G_{\theta\theta}$ by adding ${\cal A}_{\theta}^2$-term
as is usually done; this will be  discussed below.

The dual background corresponding to $SL(2)_k/U(1)_v$ with a gauge field
equal to the Lorentz connection is given by
$$
d\tilde{s}^2=k(dr^2+\coth^2 r \ d\theta^2)\ , \ \ \ \
\tilde{\phi}=\tilde{\phi_0}-  \ln\sinh r\ ,
\eq{4.3}
$$ $$
\tilde{E}
=\left(\matrix {
{\tilde G}_{\theta\theta} & -2{\tilde{\cal A}}_{\theta}\cr  0 & 1
\cr}\right) =
\left(\matrix {
k\coth^2 r & -2/\sinh^2 r \cr 0 & 1 \cr}
\right)\ .
\eq{4.4}
$$
Here $\tilde{{\cal A}}_{\theta}=\tilde{\omega}_{\theta}^r =1/\sinh^2 r$.

The background matrix $E$ and its dual $\tilde{E}$ are related by
particular $O(1,2)\subset O(2,2)$
transformations in  the following  way.
 The group $O(2,2)$ can be represented by the $4\times
4$-dimensional matrices $g$
preserving the bilinear form $J$
$$
g=\left(\matrix {a&b\cr c&d\cr} \right)\ ,\  \ \ \ \
J=\left(\matrix {0&I\cr I&0\cr} \right)\ , \eq{4.5} $$
where $a,b,c,d$ and $I$ are $2\times 2$ matrices, and
$$
g^tJg=J  \ . \eq{4.6}
$$
We define the action of $g$ on $E$ by   the fractional linear transformations:
$$
g(E)=(aE+b)(cE+d)^{-1}\ .
\eq{4.7}
 $$
The subgroup $O(1,2)\subset O(2,2)$
is generated by the elements that preserve
the heterotic structure of $E$ (namely, by those which keep the $(0,1)$ row
in $E$ invariant under (4.7)).

A particular duality element relating $E$ and $\tilde{E}$ is
$$
g_D=
\left(\matrix{ A_k&0\cr 0&A_k^{-1}\cr} \right)
\left(\matrix{ e_2&e_1\cr e_1&e_2\cr  }\right)
\left(\matrix{ A_k^{-1}&0\cr 0&A_k\cr } \right)\ ,
\eq{4.8}
   $$
i.e.
$$
g_D(E)=\tilde{E}\ ,
\eq{4.9}
$$
where
$$
A_k=\left(\matrix { \sqrt{k}&0\cr 0&1\cr } \right),\ \ \ \
e_1=\left(\matrix {1&0\cr 0&0\cr  }\right),\ \ \ \
e_2=\left(\matrix { 0&0\cr 0&1\cr } \right).\
\eq{4.10}
$$
The element
$\left(\matrix {e_2&e_1\cr e_1&e_2\cr } \right)$
in (4.8)  is called the
factorized duality in the $\theta$ direction \GMR.
It is identical to a mirror symmetry when acting on a
complex torus background matrix \GS, namely, it interchanges the
complex structure $\tau$ with its K\"{a}hler structure $\rho$.

It is also possible to relate
$E$  and $\tilde{E}$ by the $O(2,2)$ element which is found  from
$g_D$ in (4.8)
by replacing factorised duality
$\left(\matrix {e_2&e_1\cr e_1&e_2\cr } \right)$
by  the full duality transformation
$\left(\matrix {0&I\cr I&0\cr }\right)$.
This ambiguity is a result of the particular structure of the heterotic
backgrounds in (4.2),(4.4).

We now arrive  at  the important point of this section.
The duality (4.8) is not only an exact symmetry of c.f.t. and string
theory, but here it also relates one  exact heterotic string solution to
another
exact solution. Combined with  other generators of $O(1,2)$ (namely, the
generators of $GL(2)$ and $\Theta$-shifts  (i.e. constant antisymmetric tensor
background shifts)
that preserve the heterotic structure of $E$), we conclude that the full
$O(1,2)$ rotations generate exact backgrounds from exact backgrounds\foot {
The leading $\alpha'$ order
$O(d,d)$ rotations of curved backgrounds with $d$ toroidal isometries
were discussed in \Ve\GR;
in \GR\GP\ it was proven that these must correspond to exact
backgrounds.}.

An interesting question is whether
one can turn off the gauge field ${\cal A}$  with the help of
$O(1,2)$ rotations as in  refs.\GR \GP. If this were
possible,  this would correspond to   a marginal deformation which relates the
$(1,1)$ superconformal heterotic coset  solution
to  another exact  heterotic solution with ${\cal A}=0$.
 This  observation could  be useful in
order to find the exact  form of the $D=2 $ heterotic  solution  with ${\cal
A}=0$
 which has  the
perturbative expansion (2.9)--(2.11). Unfortunately, we were unable to
construct such a duality
transformation.

Let us also  note that unlike the flat case  discussed in \GRV,
here the duality acts as a fractional linear transformation on  the metric $G$
and not on  the combination $G+ \a'{\cal A}{\cal A} $  which  one  finds in
the process of
dimensional reduction.   In contrast to the  solutions discussed in \gny\GR\
here the gauge field
background  (2.3) is  $\a'$ independent (note that our $\cal A$ has canonical
dimension  $cm^{-1}$)
and hence the $\a'{\cal A}^2$-term  in $G+ \a'{\cal A}{\cal A} $   is
suppressed by an extra power of
$\a'$ or $1/k$.\foot {Note also that here $G$ is invariant under the gauge
transformations since the
latter are identified with the local Lorentz rotations  and therefore   the
$\a'{\cal A}^2$-term
is not needed for gauge invariance. }

Finally, it is obvious that the discussion in this section  can be  generalised
to the case of axial-vector duality of  heterotic  string solutions
corresponding to (1,1)
 superconformal  cosets $G/H$  for   arbitrary  group $G$  and abelian subgroup
$H$ (see
\GK).

\bigskip\bigskip
This work was supported in part by the Israel Academy of Sciences  and by the
BSF -- American-Israeli
Bi-National Science Foundation.
 A.A. Tseytlin is grateful to the Racah Institute, Jerusalem, for hospitality
while part of this  work
was done. He would like also to acknowledge a partial support of SERC.

\vfill\eject
\listrefs
\end